\documentclass[twocolumn,aps,amsmath,amssymb,showpacs]{revtex4}

\usepackage{graphicx} 
\usepackage{epsfig}
\usepackage{color}

\begin{document}

\title{Critical scaling in standard biased random walks}

\author{C. Anteneodo}
\email{celia@fis.puc-rio.br}
\author{W.A.M. Morgado}
\email{welles@fis.puc-rio.br}

\affiliation{Departamento de F\'{\i}sica, Pontif\'{\i}cia 
Universidade Cat\'olica do Rio de Janeiro,  
CP 38097, 22453-900, Rio de Janeiro, Brazil}

\begin{abstract}
The spatial coverage produced by a single discrete-time 
random walk, with asymmetric jump probability $p\neq 1/2$ and non-uniform steps, 
moving on an infinite one-dimensional lattice is investigated. 
Analytical calculations are complemented with Monte Carlo simulations. 
We show that, for appropriate step sizes, the model 
displays a critical phenomenon, at $p=p_c$. 
Its scaling properties as well as the main features of the 
fragmented coverage occurring in the vicinity of the critical point are shown. 
In particular, in the limit $p\to p_c$,  the distribution  
of fragment lengths is scale-free, with nontrivial exponents. 
Moreover, the spatial distribution of cracks (unvisited sites) defines a fractal set over the 
spanned interval. 
Thus, from the perspective of the covered territory, a very rich 
critical phenomenology is revealed in a simple one-dimensional standard model.
\end{abstract}


\maketitle


Since the beginning of the past century, random walk (RW) theory has allowed 
to deal with a diversity of problems in a number of 
areas of physics, as well as in many other theoretical and applied fields, 
e.g, biology, chemistry, computer sciences and  finance~\cite{reviews}.
The undoubtful importance of RW models, with their wide range of distinct applications,  
stems from their simplicity and effectiveness in modeling 
systems experiencing disorder, noise or randomness,  which are ubiquitous 
features of real systems. 
In particular, in physics, RWs can be seen as  
the ``harmonic oscillator'' of disordered and stochastic systems, serving 
as starting point for more realistic models.

A fundamental quantity in any phenomenon where RWs are relevant is 
the number of distinct sites visited, since it furnishes the 
extent of the active territory. 
Indeed, it is crucial in processes ranging from reaction kinetics to population 
dynamics, and also in technical applications such as in search strategies~\cite{search,cache}. 
As a consequence, analytical and numerical estimates of the covered territory 
are available for lattices of different geometry, dimensionality and  
boundary conditions~\cite{montroll,vineyard}, 
for diverse statistics of jumps, symmetric or not~\cite{2Dasym}, and other variants~\cite{varios}. 
Time covering problems~\cite{time,1Dsym} and coverage by a large number 
of RWs~\cite{Nwalks} have been investigated too. 
The vast literature on coverage mainly deals with two dimensions,   
although  there are also many works about the standard symmetric one-dimensional 
(1D) RW (e.g., \cite{montroll,vineyard,1Dsym}). Meanwhile, as far as we know,
little or no attention has been paid to the asymmetric 1D case,  
despite of its importance in biased or anisotropic processes 
such as electrophoresis, polymer translocation through pores and Brownian ratchets.
However, as we will show, the asymmetric 1D problem  presents its own 
peculiar features and nontrivial scaling properties. 

In the present work, we investigate the coverage of an infinite 1D 
regular lattice by a single RW  characterized by:   
i) asymmetry, that is, at each independent step there is 
a probability $p\neq 1/2$ to step, let us say, to the right, and additionally, ii)   
distinct step sizes in opposite directions.   
Let us call $l^+$ and $l^-$ the  sizes of the 
steps in the positive and negative directions, respectively. 
They will be expressed as integer multiples of the arbitrary lattice parameter.  
In the symmetric case $l^+=l^-$, only the positions that 
are multiple of $l^+$ are reachable. Moreover, the covered fraction of 
the interval spanned by the RW is $1/l^+$, independently of $p$.  
In particular, if $l^+=l^-=1$, complete coverage of the RW span occurs. 
However, for the asymmetric case $l^+\neq l^-$, where the two anisotropic ingredients   
compete,  a nontrivial changeover between 
different coverage regimes, dependent on $p$, may take place. 
In fact, we will show that a critical phenomenon occurs as the jump 
probability $p$ reaches a critical value. 
Moreover, we will characterize the transition as well as the partially covered, 
fragmented, states, focusing on their scaling properties.


The general basic outlines to determine the number of distinct sites visited by a RW  
can be found, for instance, in Refs.~\cite{montroll,vineyard}.
In general, the average number of different sites visited at 
step $n$, $S_n$,  can be expressed as $S_n\;=\;1+\sum_{s\neq0}\sum_{i=1}^n\,F_i(s)$,
where $F_i(s)$ is the probability that the walker 
arrives at site $s$ for the first time 
at step $i$. Moreover, $F_i(s)$ and $P_j(s)$ (the probability that, at time step $j$, 
the walker is located at integer position $s$)  are related through
\\[-6mm]
\begin{equation} \label{rel}
P_n(s)\;=\;\sum_{i=1}^n\,F_i(s)P_{n-i}(0),\;\;\;\mbox{for $n\ge1$},
\end{equation}
\\[-4mm]\noindent
while $P_o(s)=\delta_{s,0}$. 
Then, from Eq.~(\ref{rel}), one obtains the following relation between  
generating functions: $P(s,z)\;=\; \delta_{s,0}\;+\;F(s,z)P(0,z)$, 
where $P(s,z)=\sum_{n\ge 0} P_n(s)z^n$ and 
$F(s,z)=\sum_{n\ge 1} F_n(s)z^n$.
Assuming $|z|\le 1$, one obtains 
\\[-5mm]
\begin{equation}\label{sz}
S(z)\;=\;\left[ (1-z)^2\,P(0,z) \right]^{-1}, 
\end{equation}
\\[-5mm]\noindent
where $S(z)\equiv \sum_{n\ge0}S_n z^n$. 
For the present problem, it is easy to show that $P(0,z)$ explicitly is 
\\[-6mm]
\begin{equation}\label{poz}
P(0,z)\;=\;  
\sum_{k\ge0} 
\left( \begin{array}{c}
\frac{(l^++l^-)k}{l^-} \\ 
k
\end{array}\right)
\tilde{z}^\frac{(l^++l^-)k}{l^-},
\end{equation}
\\[-4mm]\noindent
with $\tilde{z}=zp^\frac{l^-}{l^++l^-}(1-p)^\frac{l^+}{l^++l^-}$.
From the definition of $S(z)$, 
the quantity $S_n$ can be obtained as $1/n!$ times the $n$th derivative of $S(z)$, 
evaluated at $z=0$.


If $l^+$ and $l^-$ have common factors, 
a mapping exists into the corresponding case of reduced (mutually prime) lengths.  
Therefore, we will restrict our study to asymmetric  
coprime couples of step lengths. 
Within the latter class of RWs, 
one has the subclass where one of the lengths is unitary. 
Let us consider as representative of this subclass, the case ($l^+,l^-$)= (2,1)  
that admits an exact solution. In this case, the sum in Eq.~(\ref{poz}) becomes
$P(0,z)\;=\;
\sum_{k\ge0} 
\left( \begin{array}{c}
3k \\ 
k
\end{array}\right)
p^k(1-p)^{2k}z^{3k}$, 
that can be reduced to
\\[-6mm]
\begin{equation}\label{poz12fin}
P(0,z)\;=\;
\Re(iy+\sqrt{1-y^2})^{1/3}/\sqrt{1-y^2}\,,
\end{equation}
\\[-6mm]\noindent
for $|y|\le 1$, where $y^2=(27/4)p(1-p)^2z^3$. 
Tauberian methods can be applied to evaluate $S_n$~\cite{montroll}. 
Alternatively, the $n$th derivative of $S(z)$ can be calculated through 
Cauchy integral formula over a suitable contour encircling the origin. 
Since $S(z)$ given by Eq.~(\ref{sz}) has one single pole in the complex plane, at $z=1$, 
then, in the limit of large $n$, one gets (after conveniently deforming the integration path)
%
$S_n=
-d\left[ z^{n+1} P(0,z)\right]^{-1}/dz|_{z=1} 
= (n+1)/P(0,1)+c_0,$
%
where $c_0$ is a constant of order 1. 
It is noteworthy that this is the same asymptotic law found for the 
standard RW, with unbiased symmetric jumps to nearest neighbors (hence $l^+$=$l^-$=1),  
but in 3D regular lattices~\cite{montroll}. 
The fraction of different sites visited (measured over the average 
length of the RW) is $f_{v,n}\equiv S_n/L_n$, where $L_n$ is 
the average total displacement. 
In the large $n$ limit, the length of the RW, for $p\neq 1/3$, 
is $L_n\sim |\langle s\rangle_n| =|3p-1|n$. 
Thus,  asymptotically, $f_{v,n}$ becomes 
$f_{v}\;=\; [|3p-1|\, P(0,1)]^{-1}$, hence,
the fraction of unvisited sites  is
\\[-6mm]
\begin{equation} \label{unvisit}
f_u=1-f_v\;=\;1-\left[ |3p-1|\,P(0,1)\right]^{-1},
\end{equation}
\\[-4mm]\noindent
where $P(0,1)$ is given by Eq.~(\ref{poz12fin}).

\begin{figure}[t!]
\centering
\includegraphics*[bb=135 300 540 540, width=0.4\textwidth]{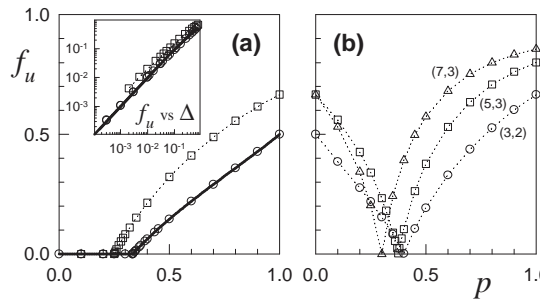}   
\caption{Fraction $f_u$ of sites left unvisited as a function of $p$. 
In all cases, symbols correspond to MC simulations. 
and  dotted lines are guides to the eye.
(a): ($l^+,l^-$)= (2,1) (circles) and (3,1) (squares). 
The full line corresponds to the theoretical prediction given by Eq.~(\ref{unvisit}). 
Inset: $f_u$ vs. $\Delta\equiv p-p_c$ in log-log scale for the same data of the 
main frame. 
(b):   ($l^+,l^-$) takes diverse coprime values indicated on the figure. 
}
\label{fig:la}
\end{figure}

Fig.~\ref{fig:la}(a) exhibits $f_u$ as a function of $p$, for ($l^+,l^-$)= (2,1). 
A transition occurs at $p_c=1/3$, where $f_u$ vanishes as $f_u=\Delta+{\cal O}(\Delta^2)$, 
with $\Delta \equiv p-p_c$, that can be derived exactly from Eq.~(\ref{unvisit}). 
For $p\le p_c$ all sites are eventually visited at least once, as expected, 
because, as soon as $\langle s \rangle_n =3\Delta<0$, 
the walker is biased towards the direction of unitary steps, 
which in turn implies  full coverage of the RW length. 
Meanwhile, for $p>p_c$, sequences of adjacent visited 
sites (fragments) are interrupted by unvisited ones. 
Therefore, the RW undergoes a transition from a fully covered state to a fragmented one.
For other instances of $(l^+,1)$, the transition occurs at the critical 
probability $p_c=1/(l^++1)$, 
where $\langle s \rangle_n=(pl^+ +p-1)n$, changes sign (driftless diffusion). 
The case ($l^+,l^-$)= (3,1), obtained by means of Monte Carlo (MC) simulations 
up to  $n\approx 10^7$ time steps, is also displayed in Fig.~\ref{fig:la}(a), 
exhibiting similar features. In both cases, $f_u(\Delta)$ 
vanishes with unitary exponent (see inset of Fig.~\ref{fig:la}(a)).

For non-unitary coprime step lengths   
(see Fig.~\ref{fig:la}(b)) a more general scenario arises. 
Full coverage occurs only at the critical point 
$p_c=l^-/(l^++l^-)$, where $\langle s\rangle_n=(pl^++(p-1)l^-)n$ is 
strictly null. 
Fragmented states are found both below and above $p_c$,  with maximal 
unvisited fractions, $f_u^-=1-1/l^-$ and $f_u^+=1-1/l^+$, respectively. 
Thus, the cases ($l^+,1$), with $l^+>1$, constitute special instances where one of the 
states is fully covered, in accordance with the fact that the 
corresponding maximal unvisited fraction $f_u^-$ vanishes.
Although we are not dealing with symmetric steps, 
notice that in the symmetric case $(1,1)$, $f_u^-=f_u^+=0$ and  
the full curve $f_u(p)$ collapses to zero, in agreement with the facts 
that there is no transition in such case and that full coverage occurs for any $p$.

\begin{figure}[ht]
\centering
\includegraphics*[bb=80 290 510 630, width=0.4\textwidth]{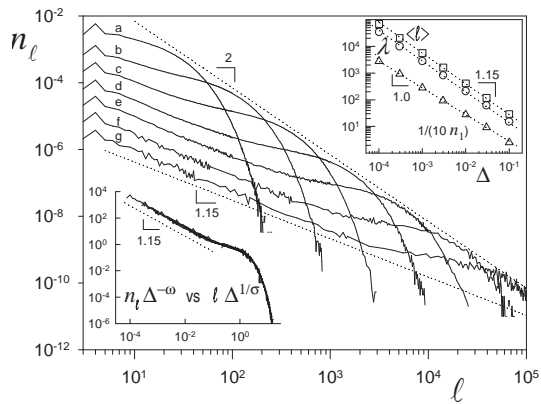}
\caption{Distribution of the sizes of covered fragments ($n_1$ was omitted),  
for ($l^+,l^-$)=(2,1)  and different 
values of $\Delta=10^{-1}$, $3\times 10^{-2}$, $\ldots$, $10^{-4}$, from a to g, 
respectively. 
Upper inset:  mean size of fragments $\langle \ell \rangle$ (squares), inverse 
exponential rate $\lambda$ (circles), and $[10n_1]^{-1}$ (triangles) as a 
function of $\Delta$. 
Lower inset:
Scaling plot of all the distributions represented in the main frame, with 
$\omega=2/\sigma$ and $1/\sigma=1.15\pm 0.05$. 
Dashed lines are drawn for comparison and their slopes indicated on the figure.   
}
\label{fig:histos}
\end{figure}

As a paradigmatic example, we will analyze 
the analytically soluble case ($l^+,l^-$)= (2,1), in the vicinity of the critical point, i.e., 
in the limit $\Delta\to 0^+$.  
In order to quantitatively characterize fragment sizes, 
the  usual computed quantities are~\cite{feder}:
\\[-4mm]
\begin{equation}
\tilde{n}_\ell  = \sum_{\ell\ge 1} n_\ell,   \;\;\;\;\;\;
\langle \ell \rangle = \sum_{\ell\ge 1} n_\ell \ell^2  
/\sum_{\ell\ge 1} n_\ell\ell  , \label{means}
\end{equation}
\\[-4mm]\noindent
where $n_\ell$ is  the mean number of fragments  
of size $\ell$, normalized per site.
Since two contiguous fragments are separated, in the (2,1) case, by one single 
unvisited site, then $\tilde{n}_\ell \approx f_u$, that 
vanishes as $\sim\Delta$ (see Fig.~\ref{fig:la}(a)).
Also, straightforwardly, $\sum_{\ell\ge 1} n_\ell\ell = 1-f_u$, that approaches 
1 in the critical limit. 
Noticing that $n_\ell\ell$ is the probability that a given site belongs 
to a fragment of size $\ell$, then, 
$\tilde{\ell} \approx \sum_{\ell\ge 1} n_\ell \ell^2$ defines the mean size 
of the fragments. 
In order to compute $\langle \ell \rangle$,  
the distribution of sizes of covered clusters (or fragments),  
$n_\ell$, was numerically built from MC simulations run up 
to $n\approx 10^6/\Delta$ steps and 
averaged over at least $10^2$ different realizations. 
The distributions for different values of $\Delta$ are displayed in Fig.~\ref{fig:histos}. 
For very large $\ell$, the decay is exponential:  $\sim \exp(-\ell/\lambda)$. 
Parameter  $\lambda$, together with $\langle \ell \rangle$, 
are plotted as a function of $\Delta$ in the upper inset of Fig.~\ref{fig:histos} 
(being $\lambda\approx \langle\ell\rangle/2\sim \Delta^{-\gamma}$, 
with $\gamma \approx 1.15$). Meanwhile, $n_1\sim\Delta$, representing a finite 
fraction of $f_u$.
In the lower inset of Fig.~\ref{fig:histos}, 
the same distributions of the main frame are scaled. 
Let us employ the standard {\em ansatz} for cluster size distributions~\cite{feder}, 
defined through: 
\\[-4mm]
\begin{equation} \label{scaling}
n_\ell(\Delta) \propto 
\Delta^\omega   \phi(\Delta^{1/\sigma} \ell)/(\Delta^{1/\sigma} \ell)^\tau  ,
\end{equation}
\\[-4mm]\noindent
where $\phi(x)$ goes to a constant value for small $x$ and   
decays exponentially in the opposite limit of large $x$. 
The power-law decay, with exponent $\tau \approx 1.15$, 
that emerges in the limit of vanishing $\Delta$ is characteristic 
of a critical behavior and signals 
the coexistence of fragments of all sizes in that limit.

By means of integral approximations to the sums 
in Eqs.~(\ref{means}) 
and employing Eq.~(\ref{scaling}), 
one gets the following relations amongst critical exponents.
Firstly, $1\approx \sum_{\ell\ge 1} n_\ell \ell \approx \int_1^\infty n_\ell \ell d\ell 
\sim \Delta^{\omega-2/\sigma}$, implying $\omega=2/\sigma$.
Secondly, $\Delta^{-\gamma} \sim \langle \ell \rangle 
 \approx \int_1^\infty n_\ell \ell^2 d\ell 
\sim \Delta^{\omega-3/\sigma}$, hence $\omega=3/\sigma-\gamma$, 
that, together with the preceding relation, implies $\gamma=1/\sigma$ and $\omega=2/\sigma$. 
The latter equality is in good accord with the behavior of the envelope of the 
distributions that has slope -2 (Fig.~\ref{fig:histos}). 
Excellent data collapse is obtained for $\omega=2/\sigma$, 
with $\gamma \approx 1.15$. 
Additionally, since $\tilde{n}_\ell \sim \Delta$, then, 
from  $\tilde{n}_\ell \approx \int_1^\infty n_\ell d\ell 
\sim \Delta^{\omega-\tau\gamma}$, it must be $\tau=2-\sigma=2-1/\gamma$. 
From the scaled histograms, we obtained $\tau\approx 1.15\pm0.05$, 
consistent with the theoretical prediction within error bars.

At this point, it is worth comparing our results with those for 
another 1D critical phenomenon, namely 1D percolation (1DP) with bonds 
connecting nearest neighbors~\cite{1Dpercolation}, 
to which many important 1D models are related~(e.g., Ref. \cite{ffire}). 
On one hand, for 1DP, $\tilde{n}_\ell \sim \Delta^{2-\alpha_p}$, 
with $\alpha_p=1$, as in the present problem. 
On the other hand, $\langle \ell \rangle\sim \Delta^{-\gamma_p}$, with $\gamma_p=1$ and 
$\omega_p=2\gamma_p=2$, values that are close but different from those found 
for the present problem.  
Moreover, the distribution of fragment sizes is a power-law, in contrast with the 
pure Poissonian one for 1DP. 
Then, we may conclude that the present model does not belong to the 1DP 
universality class.
Indeed, 
by identifying visited sites with occupied ones, the occupation probability in our 
problem is $f_v$, that tends to one in the critical limit. However, differently from 
the standard percolation problem, in the present case, unvisited sites are not 
independently located, 
e.g., if ($l^+,l^-$)=(2,1), a sequence of two or more adjacent unvisited sites 
has associated a strictly null probability of occurrence. Therefore, occupation 
correlations arise which are absent in the standard percolation problem.

\begin{figure}[h!]
\centering
\includegraphics*[bb=70 460 500 750, width=0.4\textwidth]{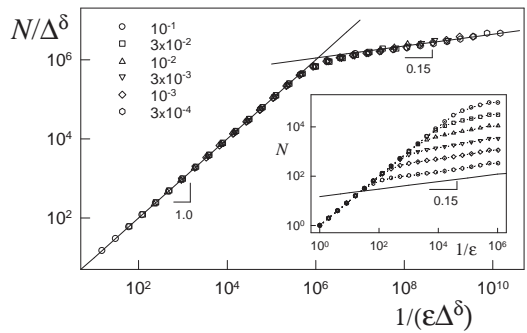}   
\caption{Scaling plot of the 
number of boxes $N$ containing unvisited sites as a function of $\varepsilon$ 
(box size in units of $L$, where $L=2^{20}$) 
for $(l^+,l^-)=(2,1)$  and different values 
of $\Delta$ indicated on the figure. 
The scaling exponent is $\delta=1/(1-d_f)$, where $d_f\simeq 0.15$. 
Inset: original plots of the data scaled in the main frame.
All solid lines are drawn for comparison and their slopes indicated on the figure.   }
\label{fig:fractal}
\end{figure}

Concerning unvisited sites, their spatial distribution was investigated through a 
box-counting procedure~\cite{feder,tel}. From the history of a single RW, 
a segment of length $L=2^{20}\approx 10^6$ was divided into 
boxes of length $2^k$, with $k\ge 0$. 
For each $\varepsilon=2^k/L$, the number of  boxes containing unvisited 
sites, $N(\varepsilon)$, was computed. 
Outcomes, accumulated over $10^2$ realizations, are displayed in Fig.~\ref{fig:fractal}. 
The neat behavior $N(\varepsilon) \sim \varepsilon^{-d_f}$, for small $\varepsilon$, 
means that the spatial distribution of 
unvisited sites constitutes a fractal set, with dimension $d_f$. 
Moreover,  the fractal exponent is in good accord with the exact
scaling relation $d_f=\tau-1$.

In summary, we have investigated the spatial coverage of single discrete-time 
anisotropic RWs, moving on an infinite one-dimensional lattice. 
Anisotropy manifests both in the length $(l^+,l^-)$, as well as in the 
probabilities $(p,1-p)$,  of jumps in opposite directions. 
We revealed the existence of a critical phenomenon in 1D that may result 
from the competition between the opposite 
trends provided by the two anisotropic ingredients (step size and step probability). 
We illustrated our findings with the particular case in which the steps are 
$(l^+,l^-)=(2,1)$, which undergoes a transition from fully to partially covered 
states as the jump probability $p$ overcomes a critical value. 
The power-law distribution of sizes of covered segments, occurring in the limit $p\to p_c^+$, 
indicates the coexistence of fragments of all lengths, with no characteristic length scale. 
Moreover, the spatial distribution of scission 
points (unvisited sites) determines a 
fractal set, in contrast with other models where the 
deposition of cracks has common statistics 
(e.g., 1D percolation~\cite{1Dpercolation}, scission model~\cite{scission}).  
It is pertinent remarking that akin features have been observed in 
one-dimensional reaction-diffusion~\cite{redif,potts}, 
$q$-state Potts spin flipping~\cite{potts} and fragmentation dynamics~\cite{fragm}, 
although criticality is attained as time evolves and critical exponents are different. 
A possible connection remains to be investigated.  

\begin{figure}[t!]
\centering
\includegraphics*[bb=140 400 560 630, width=0.4\textwidth]{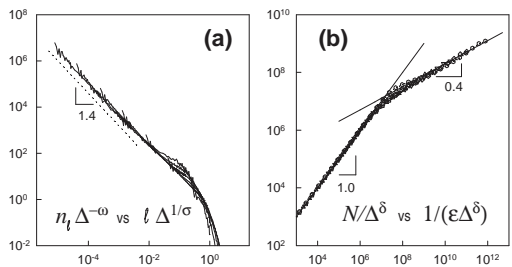}   
\caption{Critical behavior for ($l^+,l^-$)=(5,3). 
Results correspond to the limit $\Delta\to0^+$, but the same exponents 
are found in the limit $\Delta\to0^-$.
(a) Scaling plot of the distribution of the sizes of covered fragments, values of $\Delta$ 
as in Fig.~(\ref{fig:histos}), with $\omega=2/\sigma$ and $1/\sigma=1.5\pm 0.2$, 
$\tau\simeq 1.4$. 
(b) Scaling plot of the 
number of boxes $N$ containing unvisited sites as in  Fig.~(\ref{fig:fractal}). 
In this case $d_f\simeq 0.4$. 
The dotted line in (a) and solid lines in (b) are drawn for comparison and 
their slopes indicated on the figure.  
}
\label{fig:53}
\end{figure} 

Other asymmetric instances with steps ($l^+,1$), whose critical curves are  illustrated 
in Fig.~\ref{fig:la}(a), display a qualitatively similar picture 
to the case (2,1). 
Meanwhile, if both steps take non-unitary coprime values (Fig.~\ref{fig:la}(b)),  
the same critical phenomenology  
is observed in both limits $\Delta\to 0^\pm$. As a further example, scaling plots are 
also displayed, in Fig.~\ref{fig:53}, for the case ($l^+,l^-$)=(5,3) in the limit $\Delta\to0^+$.
In general, critical exponents related to the fractal dimension are not 
universal but depend on 
the step lengths, since distinct site occupation correlations take place. 

One one hand, the asymmetric RW, seen from the present perspective, 
may bear interest {\em per se} because of the nontrivial criticality contained in 
a simple model. On the other hand, it may constitute a useful statistical 
paradigm for the formation of domains or fragments by a non-equilibrium process driven 
by biased signal propagation. 
Additionally, the current coverage problem 
may be potentially useful in technical applications, e.g., 
in search strategies such as for cache hit/miss ratio optimization~\cite{cache}.

{\bf Acknoledgements:} 
We acknowledge Brazilian agencies CNPq and Faperj for partial financial support. 

\end{document}